\documentclass[aps,prd,reprint,superscriptaddress,nofootinbib]{revtex4-2}
\usepackage{amsmath}
\usepackage{mathtools}
\usepackage{amsfonts}
\usepackage[dvipsnames]{xcolor}
\usepackage{tikz}
\usepackage{graphicx}
\usepackage{hyperref}
\usepackage{braket}
\usepackage{calc}
\hypersetup{
 bookmarksnumbered=true,
 colorlinks=true,
 linkcolor=[rgb]{0.098,0.098,0.439},
 citecolor=[rgb]{0.098,0.098,0.439},
 urlcolor=[rgb]{0.098,0.098,0.439}
  }

\begin{document}

\title{
 Dynamic neuron approach to deep neural networks: Decoupling neurons for renormalization group analysis}

\author{Donghee Lee}
\email{dhlee641@kaist.ac.kr}
\author{Hye-Sung Lee}
\email{hyesung.lee@kaist.ac.kr}
\author{Jaeok Yi}
\email{wodhr1541@kaist.ac.kr}
\affiliation{Department of Physics, Korea Advanced Institute of Science and Technology, Daejeon 34141, Korea}

\date{October 2024}

\begin{abstract}
Deep neural network architectures often consist of repetitive structural elements. We introduce an approach that reveals these patterns and can be broadly applied to the study of deep learning. Similarly to how a power strip helps untangle and organize complex cable connections, this approach treats neurons as additional degrees of freedom in interactions, simplifying the structure and enhancing the intuitive understanding of interactions within deep neural networks. Furthermore, it reveals the translational symmetry of deep neural networks, which simplifies the application of the renormalization group transformation—a method that effectively analyzes the scaling behavior of the system. By utilizing translational symmetry and renormalization group transformations, we can analyze critical phenomena. This approach may open new avenues for studying deep neural networks using statistical physics.
\end{abstract}

\maketitle

\section{Introduction}

Deep learning has demonstrated remarkable performance across a diverse array of fields. Despite this success, its theoretical foundations are still in their early stages, largely due to the numerous degrees of freedom and the complexity of the deep neural network (DNN) system \cite{fan2021interpretability,li2022interpretable}.
Statistical physics is frequently used to understand complex systems with many degrees of freedom. Consequently, it is natural to expect that applying a statistical physical approach to deep learning will uncover new insights into the field.

Numerous studies have explored the connection between deep learning and statistical physics, driven by the complexity of their behaviors. Several key research areas have been thoroughly documented in the review \cite{bahri2020statistical}. Some studies address the relationship between deep learning models employing the gradient descent method and the Gibbs distribution, with the cost function acting as the Hamiltonian under specific conditions \cite{carnevali1987exhaustive, tishby1989consistent, sompolinsky1990learning, levin1990statistical, seung1992statistical, engel2001statistical, mezard2009information}. In the past, the spin glass model was proposed as a physical framework with similarities to deep learning systems, where the parameters serve as the degrees of freedom \cite{gyorgyi1990first, koebarle1990neural, mezard2009information, advani2013statistical}. Some earlier research suggested that neurons act as the degrees of freedom in deep learning systems, with Hopfield's and Little's work, as well as the Boltzmann machine, being well-known examples \cite{hopfield1982neural, little1974existence, amit1985spin, amit1987statistical, koebarle1990neural, montufar2018restricted}.

Some studies have attempted to explain the success of deep learning through the renormalization group (RG) transformation. The hierarchical structure of DNN has inspired efforts to explore the connection between DNN and RG transformation. In these studies, unimportant scaled data diminishes as it passes through layers of neurons \cite{beny2013deep, mehta2014exact, zeiler2014visualizing, lin2017does, iso2018scale, de2019deep, funai2020thermodynamics}. Additionally, the RG transformation has been suggested as a way to describe the training process of deep learning models \cite{gong2022interpreting}. An effective method for initializing DNNs from the perspective of RG has been proposed, and the scaling behavior of information is discussed in Ref.~\cite{roberts2022principles}. The RG transformation demonstrates its true power when applied to the study of phase transitions and critical phenomena in systems. Some studies have explored connections between phase transitions, critical phenomena, and deep learning \cite{gyorgyi1990first, sompolinsky1990learning, seung1992statistical, carnevali1987exhaustive,  mehta2014exact, choromanska2015loss, schoenholz2016deep, poole2016exponential, oprisa2017criticality, oprisa2017criticality1, roberts2022principles}.

Recently, intriguing phenomena have emerged in deep learning, particularly involving the (broken) neural scaling law—a principle that seems consistent across different systems. Several papers have studied the neural scaling law, with some suggesting that this behavior may be linked to phase transitions \cite{liao2020random, kaplan2020scaling, caballero2022broken, bahri2021explaining, sorscher2022beyond, hastie2022surprises, ziyin2022exact, bordelon2024dynamical, ma2024neural, du2024understanding}.

Despite significant efforts to establish a connection between deep learning and statistical physics, a concrete and general understanding of this relationship has yet to be achieved. This challenge partly stems from the complexity of the cost function. In DNNs, the iterative structure and non-linear activation functions cause the perturbative expansion of degrees of freedom—the synaptic weights and biases—to generate infinitely many non-negligible higher-order terms. In physics, solving problems of this nature often involves identifying underlying symmetries. For instance, translation symmetry plays a key role in analyzing the Ising model \cite{ernst1925beitrag}. Similarly, recognizing symmetries in a model is crucial for understanding its behavior.

In this paper, we develop a method to reformulate the cost function in a way that reveals its symmetry, thereby reducing its complexity. We refer to this as the `dynamic neuron approach' (DNA). It can be viewed as a decoupling technique that introduces additional degrees of freedom—namely, neurons in the bulk layers—coupled to the original degrees of freedom. Once these degrees of freedom have been decoupled, the system exhibits translational symmetry akin to that of the Ising model. In a specific case, we show that this approach yields a Hamiltonian resembling that of either the Ising model or the spin-glass model. Due to this symmetry, many tools from statistical physics, including RG transformations and correlation lengths, become applicable to DNNs.

The remainder of this paper is organized as follows. In Sec.~\ref{sec:Review}, we provide a brief review of the deep learning and the statistical physics approach. In Sec~\ref{sec:DynamicNeurons}, we introduce the DNA and a new Hamiltonian to analyze DNNs. In Sec.~\ref{sec:RG}, we present an example of the RG transformation using our new method. Critical phenomena are discussed in Sec.~\ref{sec:Critical}. The summary and discussion are provided in Sec.~\ref{sec:Summary}, a discussion of the spin glass model limit is included in Appendix~\ref{Appendix:SpinGlass}, and the methods for implementing the dynamic neuron system are outlined in Appendix~\ref{Appendix:Technical}.

\section{Fundamentals of Deep Learning}
\label{sec:Review}

\begin{figure}[bt]
    \centering
    \includegraphics[width=0.4\textwidth]{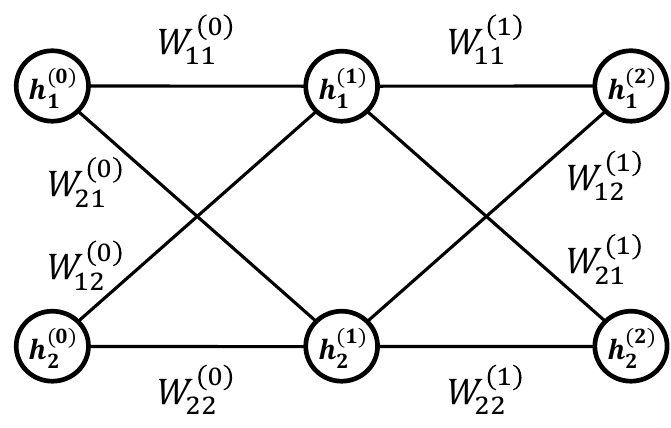}
    \caption{A DNN with $N=2$ and $M=2$, where $N$ is the width of each layer, and $M$ represents the depth of the DNN. For simplicity, the bias term $b$ has been omitted.}
    \label{fig:DNN}
\end{figure}

Before proceeding further, we overview the basic concepts of deep learning and its statistical physical approach. This section is largely based on Refs.~\cite{ruder2016overview,alzubaidi2021review,seung1992statistical}.

A DNN is a system inspired by the brain \cite{mcculloch1943logical, dongare2012introduction}. It consists of neurons, synaptic weights, and synaptic biases, with neurons organized into layers. The $i$th neuron in the $(m+1)$th layer, denoted as $h_i^{(m+1)}$, is influenced by all neurons in the $m$th layer, which are connected via synaptic weights $W_{ij}^{(m)}$ and synaptic biases $b_i^{(m)}$. (See Fig.~\ref{fig:DNN}.) This relationship is described by the following equation:
\begin{equation}\begin{split}
h_i^{(m+1)}& = \sigma\Big(\sum_j W_{ij}^{(m)} h_j^{(m)} + b_i^{(m)}\Big)\\
&\equiv\sigma^{(m)}_i. \label{eq: recursive relation}
\end{split}\end{equation}
In this formula, $\sigma$ represents a nonlinear function known as the activation function.\footnote{Typically, the sigmoid function or rectified linear unit(ReLU) is used as an activation function. An example of the sigmoid function is $\sigma(x) = \frac{1}{1 + e^{-x}}$, and for ReLU it is $\sigma(x) = \max(x, 0)$.} The repetitive calculation of Eq.~\eqref{eq: recursive relation} determines the value of each neuron, starting from the input (initial layer) $x = h^{(0)}$ and eventually reaching the output (final layer) $z = h^{(M)}$.

Simply put, deep learning is a method for approximating an unknown function using a DNN. The process of building a neural network from a given dataset is called training. 
There are several types of deep learning classified by their training methods, but we will focus on supervised learning for the sake of simplicity. In supervised learning, the machine is trained to give a good approximation for a set of examples, known as a training dataset. The training dataset consists of $L$ data vectors $x^{[l]}$, $l = 1, 2, \cdots, L$, and their corresponding label vectors $y^{[l]}$. For simplicity, we will often omit $[l]$, provided it does not cause confusion.

The goal of the training process is to minimize the discrepancy between the label $y^{[l]}$ and the output of the DNN $z^{[l]}$ by adjusting the synaptic weights $W_{ij}^{(m)}$ and biases $b_i^{(m)}$. 

The discrepancy between the label and the output is typically measured using the $L^2$- or $L^1$-norm, referred to as the cost function or training error. The cost function using the $L^2$-norm is given as\footnote{Although it is customary to define the cost function by dividing Eq.~\eqref{eq:costFunction} by the number of data points $L$, we do not follow this convention. Since $L$ can be absorbed into the step size, it does not affect our discussion.}
\begin{equation}
\begin{split}
    C=\sum_{i,l}\left(y_i^{[l]}-z_i^{[l]}\right)^2.
\end{split}
\label{eq:costFunction}
\end{equation}

A popular method for reducing the cost function is gradient descent. In each step of adjusting the synaptic weights and biases, the gradient of the cost function is calculated, and the synaptic weights and biases are adjusted by subtracting this value:
\begin{equation}\begin{split} \label{eq:gradient_descent}
    \Delta W_{ij}^{(m)}=-\eta\frac{\partial C}{\partial W_{ij}^{(m)}}, \qquad     \Delta b_{i}^{(m)}=-\eta\frac{\partial C}{\partial b_{i}^{(m)}},
\end{split}\end{equation}
where the step size $\eta$ determines the magnitude of the weight and bias updates. These equations show that the DNN is constructed by determining the values of the weights and biases, not the neurons. In other words, only the weights and biases are the degrees of freedom in the system, while the neurons serve as placeholders in Eq.~\eqref{eq: recursive relation}.

After the training of a DNN is complete, it is used for general data; this process is called inference. The main purpose of deep learning is to train a DNN that makes good inferences. 
Addressing inference presents many challenges, which we plan to discuss in future work.

From a physical perspective, Eq.~\eqref{eq:gradient_descent} represents the equation of motion for the system.\footnote{In Eq.~\eqref{eq:gradient_descent}, the step size can be interpreted as the time interval between weight updates. Since it does not influence the underlying physics, we set $\eta = 1$.} To better capture the stochastic nature of training in deep learning, one may consider a Langevin-type equation of motion \cite{AMATO1991207, BURTON1992627, welling2011a,chen2014stochastic, Li_Chen_Carlson_Carin_2016}, where the update of weights includes both deterministic and stochastic components:
\begin{equation} \frac{dw}{dt} = -\nabla_w C + \eta(t), \end{equation}
where $\eta(t)$ represents a white Gaussian noise with zero mean and time correlation $\langle \eta(t)\eta(t')\rangle\propto T \delta_{ij}\delta(t-t')$. Here, $T$ plays the role of an effective temperature that quantifies the strength of fluctuations in the system.

It is known that this equation of motion drives the system to a Gibbs distribution \cite{seung1992statistical, Risken1996, PhysRevX.11.031059, Pavliotis2014}. Specifically, after a sufficiently long period of training, the probability distribution for a particular synaptic configuration ${W, b}$ is given by:
\begin{equation}\begin{split}\label{eq:probability} P(W,b) = \frac{1}{Z} e^{-\beta C}, \end{split}\end{equation}
where $\beta = \frac{1}{T}$, and the temperature $T$ arises from the strength of the stochastic noise in the training dynamics, as described above. In this paper, we adopt a framework wherein the synaptic weights of the DNN follow a Gibbs distribution, with the cost function serving as the energy. The partition function, denoted by $Z$, is defined as:
\begin{equation}\begin{split} Z \equiv \sum_{\{W,b\}} e^{-\beta C}, \end{split}\end{equation}
where $\{W, b\}$ represents all possible configurations of synaptic weights and biases.\footnote{Considering all configurations of the system implies accounting for an infinite amount of time in the training process. However, in most examples from statistical physics, it is widely observed that a sufficiently long experimental run ensures that the system follows the Gibbs distribution. Therefore, if the training time is sufficiently long, we expect our argument to apply to such a system.}

Any expectation value of observables can be calculated as follows: \begin{align}\langle O \rangle = \frac{1}{Z} \sum_{\{W,b\}} O e^{-\beta C}.\end{align} As one can see, the cost function $C$ plays the role of the Hamiltonian. Therefore, it will be denoted as $H$ from now on.

\section{Dynamic neurons}
\label{sec:DynamicNeurons}

As mentioned earlier, during training, the change in synaptic weights $W_{ij}^{(m)}$ and biases $b_i^{(m)}$ over time follows the equation of motion defined by the gradient descent. This indicates that synaptic weights and biases represent the system's degrees of freedom. The cost function, or Hamiltonian, governs the probability distribution of degrees of freedom after the training process.
\begin{equation}
    H_\text{original} = \sum_{i,l} \left(y_i^{[l]} - z_i^{[l]} \right)^2 \label{eq:original H}
\end{equation}
where
\begin{equation}
\begin{split}
    z_i^{[l]}& =\sigma \Big(\sum_{i_1} W_{ii_1}^{(M)} \sigma\Big(\sum_{i_2} W_{i_1 i_2}^{(M-1)}\\& \hspace{7mm} \cdots \sigma\Big(\sum_{i_M} W_{i_{M-1} i_M}^{(0)} x_{i_k}^{[l]} + b_i^{(0)}\Big)+b_i^{(1)}\Big)\cdots\Big).
    \label{eq:iterative structure}
    \end{split}
\end{equation}

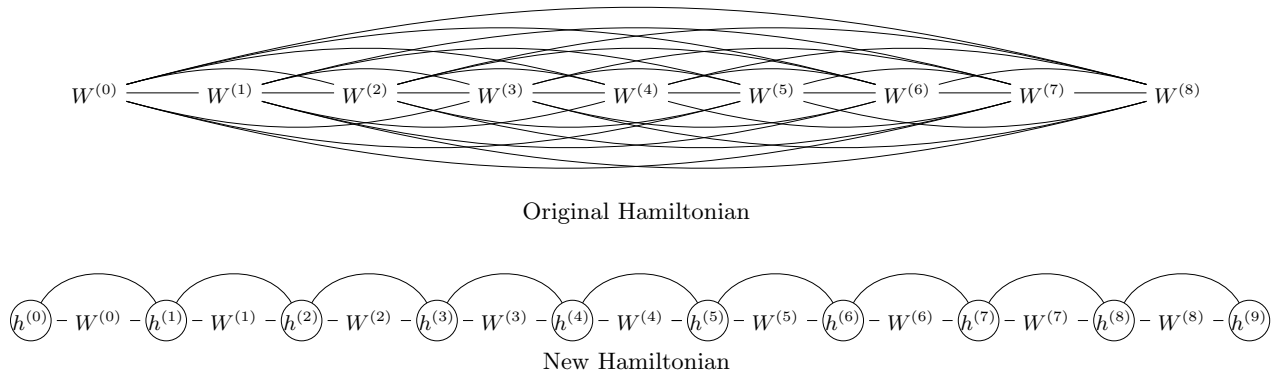
\begin{figure*}
\scalebox{.9}{\begin{tikzpicture}
    \node at (0,0) (0) {$W^{(0)}$};
    \node at (2,0) (1) {$W^{(1)}$};
    \node at (4,0) (2) {$W^{(2)}$};
    \node at (6,0) (3) {$W^{(3)}$};
    \node at (8,0) (4) {$W^{(4)}$};
    \node at (10,0) (5) {$W^{(5)}$};
    \node at (12,0) (6) {$W^{(6)}$};
    \node at (14,0) (7) {$W^{(7)}$};
    \node at (16,0) (8) {$W^{(8)}$};
    \draw (0) -- (1);\draw (1) -- (2);
    \draw (2) -- (3);\draw (3) -- (4);
    \draw (4) -- (5);\draw (5) -- (6); \draw (6) -- (7);\draw (7) -- (8);
    \draw (0)  to[bend left=15] (2);
        \draw (1)  to[bend left=15] (3);    \draw (2)  to[bend left=15] (4);
        \draw (3)  to[bend left=15] (5);
        \draw (4)  to[bend left=15] (6);
        \draw (5)  to[bend left=15] (7);
        \draw (6)  to[bend left=15] (8);
    \draw (0)  to[bend right=15] (3);
        \draw (1)  to[bend right=15] (4);    \draw (2)  to[bend right=15] (5);
        \draw (3)  to[bend right=15] (6);   \draw (4)  to[bend right=15] (7);
        \draw (5)  to[bend right=15] (8);
    \draw (0)  to[bend left=15] (4);
        \draw (1)  to[bend left=15] (5);    \draw (2)  to[bend left=15] (6);
        \draw (3)  to[bend left=15] (7);    \draw (4)  to[bend left=15] (8);
        \draw (0)  to[bend right=15] (5);
        \draw (1)  to[bend right=15] (6);
        \draw (2)  to[bend right=15] (7);
        \draw (3)  to[bend right=15] (8);
        \draw (0)  to[bend left=15] (6);
        \draw (1)  to[bend left=15] (7);
        \draw (2)  to[bend left=15] (8);
        \draw (0)  to[bend right=15] (7);
        \draw (1)  to[bend right=15] (8);
        \draw (0)  to[bend left=15] (8);
\end{tikzpicture}}

Original Hamiltonian

\vspace{5mm}
\scalebox{.9}{
\begin{tikzpicture}
    \node at (0,0) (0) {$W^{(0)}$};
    \node at (2,0) (2) {$W^{(1)}$};
    \node at (4,0) (4) {$W^{(2)}$};
    \node at (6,0) (6) {$W^{(3)}$};
    \node at (8,0) (8) {$W^{(4)}$};
    \node at (10,0) (10) {$W^{(5)}$};
    \node at (12,0) (12) {$W^{(6)}$};
    \node at (14,0) (14) {$W^{(7)}$};
    \node at (16,0) (16) {$W^{(8)}$};
    \node at (-2+1,0) (-1) {$h^{(0)}$};
    \node at (0+1,0) (1) {$h^{(1)}$};
    \node at (2+1,0) (3) {$h^{(2)}$};
    \node at (4+1,0) (5) {$h^{(3)}$};
    \node at (6+1,0) (7) {$h^{(4)}$};
    \node at (8+1,0) (9) {$h^{(5)}$};
    \node at (10+1,0) (11) {$h^{(6)}$};
    \node at (12+1,0) (13) {$h^{(7)}$};
    \node at (14+1,0) (15) {$h^{(8)}$};
    \node at (16+1,0) (17) {$h^{(9)}$};
    \draw (-1) circle (.3);
    \draw (1) circle (.3);
    \draw (3) circle (.3);
    \draw (5) circle (.3);
    \draw (7) circle (.3);
    \draw (9) circle (.3);
    \draw (11) circle (.3);
    \draw (13) circle (.3);
    \draw (15) circle (.3);
    \draw (17) circle (.3);
    \draw (-1) -- (0);
    \draw (0) -- (1);    \draw (1) -- (2);
    \draw (2) -- (3);    \draw (3) -- (4);
    \draw (4) -- (5);\draw (5) -- (6);\draw (6) -- (7);\draw (7) -- (8);\draw (8) -- (9);\draw (9) -- (10);  \draw (10) -- (11);  \draw (11) -- (12); \draw (12) -- (13);   \draw (13) -- (14);   \draw (14) -- (15);    \draw (15) -- (16); \draw (16) -- (17);
    \draw (-1) to[bend left=60] (1);    \draw (1) to[bend left=60] (3);   \draw (3) to[bend left=60] (5);\draw (5) to[bend left=60] (7);\draw (7) to[bend left=60] (9);\draw (8) -- (9);\draw (9) to[bend left=60] (11);    \draw (11) to[bend left=60] (13); \draw (13) to[bend left=60] (15); \draw (15) to[bend left=60] (17);
\end{tikzpicture}}

New Hamiltonian
\caption{Interactions in the original and new Hamiltonians. For simplicity, each layer includes only one neuron, and synaptic biases are omitted. The upper figure depicts the original Hamiltonian, while the lower figure illustrates the new Hamiltonian. Degrees of freedom are connected if a direct coupling exists between them. (Self-interactions are omitted in both cases.) Circles in the lower figure represent dynamic neurons, the newly introduced degrees of freedom.}
\label{fig:1dinteraction}
\end{figure*}

In this Hamiltonian, the synaptic weights and biases are iteratively plugged into the activation function, as specified by Eq.~\eqref{eq: recursive relation}. Although DNNs are usually described as shown in Fig.~\ref{fig:DNN}, this iterative structure reveals highly complex interactions between degrees of freedom, as depicted in the upper figure of Fig.~\ref{fig:1dinteraction}. This complexity can be understood by considering the Taylor series expansion of the activation function, which involves complicated products of synaptic weights and biases. In particular, even the distant degrees of freedom can interact. Additionally, synaptic weights and biases from different layers contribute differently to $H$, increasing the asymmetry and complicating the analysis further. To apply various concepts from statistical physics, we need to compute the partition function through the following integral:
\begin{equation} Z = \int dW db \exp[-\beta H_\text{original}]. \end{equation}
This integration is challenging to compute, as the Hamiltonian consists of intricate terms involving synaptic weights and biases. Therefore, managing the Hamiltonian and the partition function in their original forms is immensely difficult.

To address this difficulty, we promote the neurons $h_i^{(m)[l]}$ to additional dynamic degrees of freedom and ensure that Eq.~\eqref{eq: recursive relation} is dynamically satisfied. Instead of using iterated activation functions, we introduce a series of Dirac delta functions and simplify the exponent to the absolute square as follows:
\begin{equation}
\begin{split}
Z = \int dW db dh \prod_{i,l} \bigg[
& \exp\Big[- \beta \left(y^{[l]}_i-h^{(M)[l]}_i\right)^2\Big]\\
& \hspace{2mm}\prod_{m}\delta\left(h_i^{(m+1)[l]}-\sigma_i^{(m)[l]}\right)  \bigg].
\end{split}
\end{equation}
Here, $i$ runs over the widths, $l$ runs over the data numbers, and $m$ runs over the layers.
We refer to these new dynamic degrees of freedom, $h^{(m)[l]}$, as ``dynamic neurons.'' While this partition function appears straightforward, it does not yet simplify the computation.

To further evaluate the calculations, we employ an approximation of the Dirac delta function\footnote{Despite the name, the Dirac delta function is not a function but a distribution or generalized function. This means that the approximation may not be strictly valid in the rigorous sense, though it is frequently employed in the literature. We adopt this approximation because the partition function involves integration, and we expect it to yield accurate results once integrated.} 
as the limit of a Gaussian distribution:
\begin{equation}\label{eq: Dirac delta approximation} \delta(z) \approx \lim_{v \rightarrow 0^+}\frac{1}{v \sqrt{2\pi}} \exp\left[- \frac{z^2}{2v^2} \right]. \end{equation}

Applying this approximation, the partition function becomes:
\begin{equation}
\begin{split} 
    Z=\int  d & W db  dh \lim_{v\to 0^+}\left(\frac{1}{v\sqrt{2\pi }}\right)^{n}\\  & 
\exp\Big[-\beta \sum_{i,l} \left(y_i^{[l]}-\sigma^{(M)[l]}_i\right)^2 \\ &\qquad \quad 
- \beta_v \sum_{i,l,m}\left(h^{(m+1)[l]}_i-\sigma_{i}^{(m)[l]}\right)^2 \Big] ,\label{eq:modified pf}
\end{split}
\end{equation}
where $\beta_v = 1/2v^2$.\footnote{Although distinct values of $v$ could be assigned to each term, we assume $v$ remains the same for all terms. If a limit exists as multiple distinct values of $v$ approach 0, the same limit will be achieved when the values of $v$ are set identically and approach 0. This can also be understood through the concept of a Lagrange multiplier. If we interpret $\beta_v$ as the Lagrange multiplier, it imposes the constraint that the sum of the absolute square terms must vanish. For this to occur, each individual term must also vanish.} 

We observe that incorporating the constraint terms reveals the internal structure of a DNN within the Hamiltonian framework. By defining the new Hamiltonian\footnote{If the cost function is defined using an $L^p$-norm, we can define the new Hamiltonian as $H_\text{new} = \sum \lambda |h -\sigma|^p$.
}  as \begin{equation}\label{eq: modified Hamiltonian} H_\text{new} = \sum_{i,l,m} \lambda_m \left(h_i^{(m+1)[l]}- \sigma_i^{(m)[l]} \right)^2 \end{equation} with $\lambda_m = \beta_v / \beta$ except for $\lambda_M = 1$, we can denote Eq.~\eqref{eq:modified pf} in the typical way of describing the partition function,
\begin{equation}
    Z = \int dW db dh \exp[-\beta H_\text{new}].
\end{equation}
where the limit over $v$ is omitted for simplicity. The form of this new Hamiltonian is the summation of interactions between the nearest layers [$m$-th layer and $(m+1)$-th layer]. In other words, interactions in the new Hamiltonian are restricted to the nearest degrees of freedom following the same pattern, unlike the original Hamiltonian, thereby implying locality.

Physically, this approach can be viewed as a decoupling strategy. Dynamic neurons are introduced between the original degrees of freedom, effectively simplifying the complex interactions between weights and biases. These interactions are decoupled and replaced with simpler interactions between dynamic neurons and their nearest weights and biases, analogous to using power strips to organize and untangle cables. The structure of the new Hamiltonian is illustrated in the lower part of Fig.~\ref{fig:1dinteraction}, where the original Hamiltonian's complex structure shown in the upper figure is decoupled.

Although we introduced the infinite $\beta_v$ in the previous discussion, we will now study a more general system with an arbitrary positive $\beta_v$. When $\beta_v$ is sufficiently large, the system is expected to behave similarly to the infinite $\beta_v$ case. In other words, we are considering a generalization of deep learning in which the limit of large $\beta_v$ corresponds to the typical DNN.

Since Eq.~\eqref{eq: modified Hamiltonian} is written as a sum of terms with the same form, the Hamiltonian is invariant under translation,
\begin{equation}
h_i^{(m)[l]}\rightarrow h_i^{(m+1)[l]}, \quad \sigma_i^{(m)[l]}\rightarrow \sigma_i^{(m+1)[l]}
\end{equation}
except at the boundaries (the first and last layers). This approximate symmetry is not easily identifiable in the original Hamiltonian given by Eqs.~\eqref{eq:original H} and \eqref{eq:iterative structure}.

The emergence of this type of translation symmetry can simplify calculations and lead to interesting consequences. For example, in the one-dimensional and two-dimensional Ising models, the transfer matrix can be calculated explicitly due to translational symmetry. Therefore, it is expected that the new Hamiltonian can be analyzed by utilizing this symmetry. As an example, we perform the RG transformation in Sec.~\ref{sec:RG}.

This symmetry is approximate due to the system's boundaries. However, it is generally assumed that as the system size increases, the symmetry becomes more precise. Similarly, in a DNN with considerable depth, the symmetry becomes nearly exact. Since depth is regarded as a key factor in the success of deep learning \cite{hinton2006reducing, krizhevsky2012imagenet, hinton2012deep, ciregan2012multi, sun2016depth}, this approximate symmetry may offer insights into the role of depth in understanding neural networks.

The new Hamiltonian in Eq.~\eqref{eq: modified Hamiltonian} is expressed as the sum of the original cost function (for $m = M$) and additional constraint terms, each multiplied by a constant $\lambda_m$ (for $m \neq M$).This approach is analogous to the use of Lagrange multipliers in Lagrangian mechanics. Notably, similar terms arise when the Yang-Mills Lagrangian is quantized using the Fadeev-Popov procedure in quantum field theory \cite{faddeev1967feynman}, although we will not discuss this further here.

Moreover, this system can be effectively described using principles from statistical physics, which are often employed to uncover the macroscopic behavior of complex systems composed of similarly interacting units.

With certain simplifying assumptions, this approach yields a Hamiltonian that closely resembles the spin glass model, a well-established framework in statistical physics.\footnote{Several approaches have already introduced spin glass models to analyze DNNs under various assumptions \cite{gyorgyi1990first,koebarle1990neural, mezard2009information, advani2013statistical, hopfield1982neural, little1974existence, amit1985spin, amit1987statistical}.} (See Appendix \ref{Appendix:SpinGlass}.) This approach offers a new method for applying the spin glass model to gradient descent-based training in DNNs. Also, since we represented the training process of a general DNN using tools from statistical physics, we can inversely implement a system governed by such a Hamiltonian through the training of a DNN. (See Appendix~\ref{Appendix:Technical}.)

\section{Renormalization Group Transformation}
\label{sec:RG}

Since the new Hamiltonian resembles well-known statistical models, we can apply statistical physics tools to analyze the training of DNNs. A promising method in statistical physics is the RG transformation, which is widely used in fields such as condensed matter and high-energy physics. This technique systematically simplifies a system by progressively integrating out smaller-scale fluctuations, leading to a renormalized system with fewer degrees of freedom.

\begin{figure}[b]
    \centering
    \includegraphics[width=0.8\linewidth]{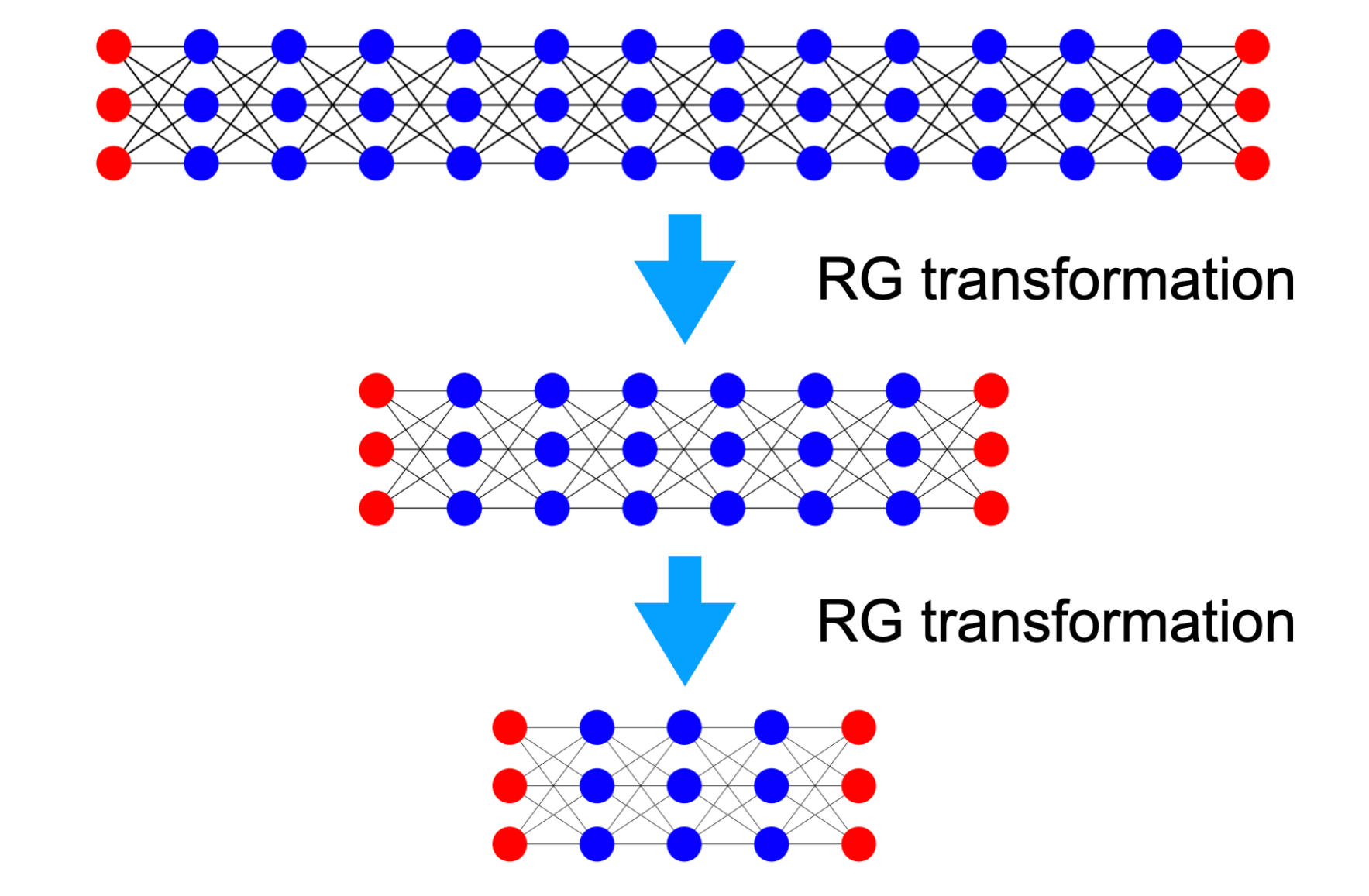}
    \caption{RG transformation applied to the DNN using the DNA, resulting in reduced network depth. A notable similarity with the Ising model is apparent.
    }
    \label{fig:RGtransformation}
\end{figure}

To understand how the components of the DNN function, we apply the RG transformation with respect to depth using our new Hamiltonian, as shown in Fig.~\ref{fig:RGtransformation}.\footnote{After an RG transformation, the system's degrees of freedom are reduced, making the Hamiltonian primarily distance-dependent along the depth direction. This complicates applying the RG transformation with respect to the width, as maintaining the Hamiltonian's form during such a transformation is challenging.} We begin by examining a simple scenario under specific assumptions. While these assumptions may oversimplify the system, potentially differing from actual DNNs, this example is designed to illustrate how the RG transformation can be applied to DNN within the context of the new Hamiltonian.

(i) The system consists of one neuron per layer ($N=1$) with a large depth ($M$ is large), and the training dataset contains a single data point ($L=1$). This assumption may be oversimplification and the example could be quite far from actual DNNs. In fact, the Modified National Institute of Standards and Technology (MNIST) \cite{lecun1998gradient} data consist of 60000 training examples ($L=60000$), and the universal approximation theorem prefers a sizable width with $N > 1$ \cite{lu2017expressive,hanin1710approximating, johnson2018deep, kidger2020universal, park2020minimum}. A lack of data can lead to insufficient training and overfitting, while insufficient depth may undermine the validity of the universal approximation theorem. It should be highlighted that our approach can be generally applied to DNNs with arbitrary $N,\  M,\  L$. Nevertheless, we make this assumption because our primary interest here lies in analytically demonstrating the statistical physics properties.

(ii) The activation function is the Heaviside step function:
\begin{equation}
    \sigma(x)=\begin{cases}
        0 & x<0 \\
        1 & \text{otherwise}
    \end{cases}
\end{equation}
Although the Heaviside step function is rarely used in modern DNNs due to its zero gradient, it was employed in earlier models and applications \cite{mcculloch1943logical, sharma2017activation}.

(iii) The weights $W$ and biases $b$ are restricted to values of $-1$ or $1$, and the neurons $h$ are limited to values of $0$ or $1$. While, in theory, $h$ should be integrated over all possible values, we expect the Gaussian-like behavior of the partition function to constrain $h$ to values near $0$ or $1$. Although in general DNNs $W$ and $b$ can take arbitrary real numbers, we restrict their values here for simplicity. Interestingly, recent research has explored neural networks where parameters are similarly restricted to discrete values, such as $-1, 1$ or $-1, 0, 1$. Despite these limitations, that model appears to achieve good performance \cite{wang2023bitnet, ma2024era}.

Given the restrictions on the available values, integration in the partition function is replaced by summation. Thus, the partition function for this example is written as:
\begin{equation}
    Z = \sum_{\{W, b, h\}} \exp[-\beta H]
\end{equation}
where the summation is over all possible values of $W$, $b$, and $h$.
Since we are considering the case with $N=1$ and $L=1$, the new Hamiltonian in Eq.~\eqref{eq: modified Hamiltonian} can be expressed as a sum:
\begin{equation}
    H = \sum_m \lambda_m H_m \label{eq:sum}
\end{equation}
where each $H_m$ has the common form:
\begin{equation}
    H_m = \left[h^{(m+1)} - \sigma\left(W^{(m)} h^{(m)} +b^{(m)}\right)\right]^2 .
\end{equation}

As discussed in Sec.~\ref{sec:DynamicNeurons}, the Hamiltonian exhibits translational symmetry similar to that of the Ising model. Consequently, we can apply the RG transformation in a manner similar to its application in the Ising model. One approach involves summing over $W^{(m)}$, $b^{(m)}$, and $h^{(m)}$ for even-numbered layers in the partition function. After performing this summation, the partition function describes a system with fewer degrees of freedom. This real-space RG transformation method is known as decimation \cite{hu1982introduction, cardy1996scaling}, and the translational invariance or periodicity plays a crucial role in performing the decimation.\footnote{Decimation is generally avoided for systems with dimensions higher than one because it fails to account for the renormalization of the magnitude of the degrees of freedom, preventing it from achieving a true fixed point in higher dimensions.}

Extracting the terms related to $W^{(m)}$, $b^{(m)}$, and $h^{(m)}$ from the partition function, we have
\begin{equation}
    Z_m = \sum_{\{W^{(m)}, b^{(m)}, h^{(m)}\}} \exp[-\beta H_{m+1} - \beta H_m] .
\end{equation}
This summation is finite and can be calculated without difficulty:
\begin{equation}
    Z_m=3u^{-2}u^{2 h+2 \sigma}+ u^{-1}(u^{2\sigma}+2u^{2h})+2
\end{equation}
where
\begin{equation}
\begin{split}
    &h = h^{(m+1)} ,\\
    &\sigma = \sigma(W^{(m-1)}h^{(m-1)} + b^{(m-1)}) ,\\
    &u=e^{\beta_v} .
\end{split}
\end{equation}
By taking the logarithm, we obtain the Hamiltonian of the renormalized system, denoted as $H'$:
\begin{equation}
    -\beta H_m'=\ln[3u^{-2}u^{2 h+2 \sigma}
    + u^{-1}(u^{2\sigma}+2u^{2h})+2] .
\end{equation}
By making use of the restricted configuration, where $h^2 = h$ and $\sigma^2 = \sigma$, we can express the renormalized Hamiltonian in polynomial form:\footnote{This approach of representing complex functions as polynomials is also observed in the RG transformation of the Ising model.}
\begin{equation}
\begin{split}
    -\beta H_m'
    &=h \sigma\ln(3u^{2}+3u+2)\\
    & \quad +(1-h)\sigma\ln(u+2u^{-1}+5)\\
    &\quad +h(1-\sigma)\ln(2u+u^{-1}+5)\\
    &\quad +(1-h)(1-\sigma)\ln(3u^{-2}+3u^{-1}+2)\\
    &=h\sigma \ln\frac{(3u^2+3u+2)(2u^2+3u+3)}{(u^2+5u+2)(2u^2+5u+1)}\\
    &\quad +h^2\ln\frac{u(2u^2+5u+1)}{2u^2+3u+3}\\
    &\quad +\sigma^2 \ln\frac{u(u^2+5u+2)}{2u^2+3u+3}\\
    &\quad +\text{(constant)} .
    \end{split}
\end{equation}

In general, suppose that our Hamiltonian takes the following form:
\begin{align}
    -\beta H&=\sum_m  \left(Ah\sigma+Bh+C\sigma+D\right) .
\end{align}
The coupling constant $A$ has two critical aspects worth emphasizing. First, it introduces the interaction between neurons across different layers, effectively encoding the mechanism of information propagation within the system. Consequently, any changes in $A$ under RG transformation are expected to significantly influence the learning dynamics and overall behavior of the deep learning model. Second, this coupling term bears resemblance to the spin-spin interaction term in the Ising model. In the Ising model, the spin-spin interaction is the primary driver behind phase transitions, reflecting collective behavior changes at critical points. Similarly, in the context of the dynamic neuron system, $A$ is likely to play a pivotal role in phase transitions, potentially governing shifts in the system’s operational regimes or learning efficiency. Understanding this analogy provides a deeper insight into how critical phenomena might emerge in such neural systems.

\begin{figure}[t]
    \centering
    \includegraphics[width=0.8\linewidth]{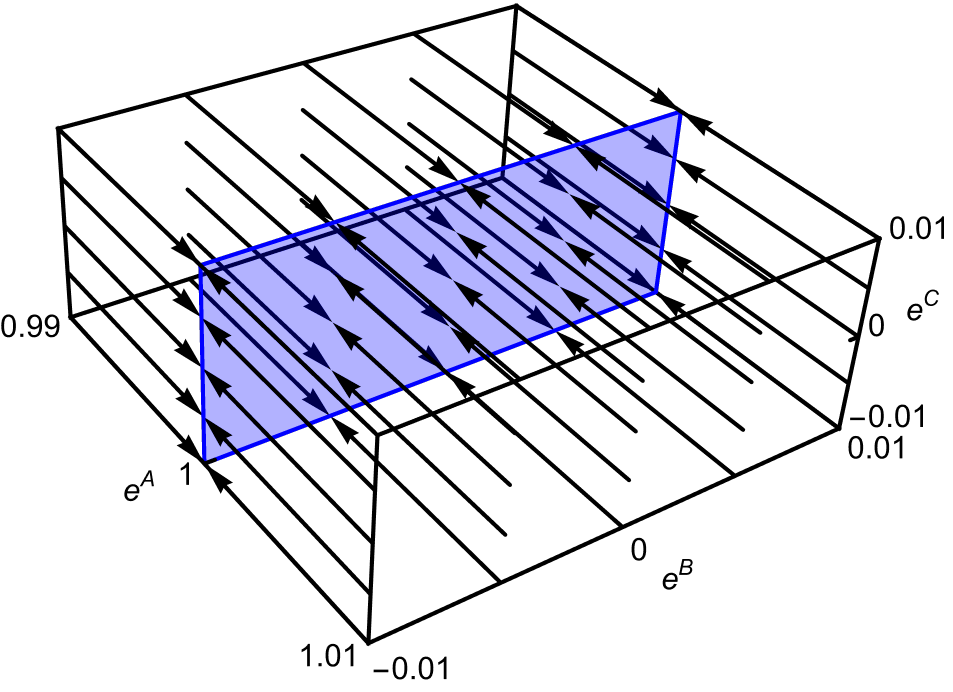}
    \caption{RG flow near the fixed points $e^{A^*}=1$, demonstrating that RG flows converge to the plane composed of fixed points.}
    \label{fig:RG flow}
\end{figure}

Under the RG transformation, the constant term in the Hamiltonian only affects the non-singular part of the free energy and is typically ignored when calculating the critical properties of the system.

\begin{equation}
    \begin{split}
        A'&=\ln \left[\frac{2+2e^{A+C}+e^{A+B}+3e^{2A+B+C}}{2+2e^{A+C}+e^B+3e^{A+B+C}}\right. \\ & \left. \qquad \qquad \quad \qquad \frac{2+2e^C+e^B+3e^{B+C}}{2+2e^{C}+e^{A+B}+3e^{A+B+C}} \right]\\
        B'&=B+\ln \frac{2+2e^{A+C}+e^B+3e^{A+B+C}}{2+2e^C+e^B+3e^{B+C}}\\
        C'&=C+\ln \frac{2+2e^{C}+e^{A+B}+3e^{A+B+C}}{2+2e^C+e^B+3e^{B+C}}\\
        \end{split}
\end{equation}

The fixed points are the points that the renormalized system is equal to itself.
By obtaining the solutions $(A,\ B,\ C)$ of the equation
\begin{equation}
    (e^{A'}, e^{B'}, e^{C'}) = (e^{A}, e^{B} , e^{C} ),
\end{equation}
the fixed points $(A^*,\ B^*, \ C^*)$ of this system can be found. The exponents are used to simplify the calculations. The fixed points are given as
\begin{equation}
  e^{A^*} = 1 \label{eq:fpt}
\end{equation}
for arbitrary values of $e^{B^*}$ and $e^{C^*}$. 
Figure~\ref{fig:RG flow} illustrates the RG flows near these fixed points. The RG flows converge to the fixed points, indicating that there is no phase transition.

\begin{figure}[t]
    \centering
    \includegraphics[width=0.8\linewidth]{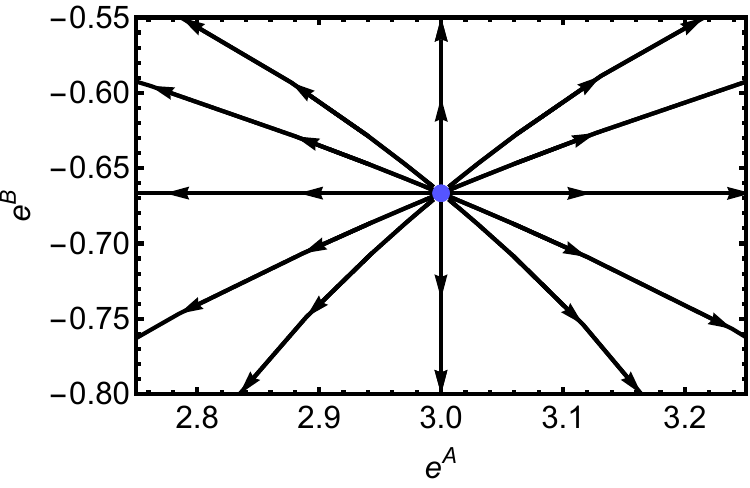}
    \caption{Diverging RG flow near the fictitious fixed point $(e^{A^*},e^{B^*},e^{C^*})=(3,-2/3, -1/3)$. For visibility, $e^C = -1/3$ is fixed for all flows.}
    \label{fig:fictitious}
\end{figure}

Since the couplings are real, there are no fixed points other than those given in Eq.~\eqref{eq:fpt}. However, if we assume that $e^{A^*}$, $e^{B^*}$, and $e^{C^*}$ can be negative, we can find another fixed point at
\begin{equation}
e^{A^*} = 3, \ e^{B^*} = -\frac23 , \ e^{C^*} = - \frac13 .
\end{equation}
The RG flow near this fictitious fixed point is shown in Fig.~\ref{fig:fictitious}, where we observe that the RG flows diverge from this point. This fixed point appears unphysical since it can only be achieved when the couplings are imaginary. Nevertheless, we anticipate that more general and complex deep learning models, beyond our simplifying assumptions (i)–(iii), will exhibit unstable fixed points, or critical points, where RG flows diverge within realistic coupling regions.\footnote{This reasoning is grounded in insights from statistical physics. For instance, while the one-dimensional Ising model lacks a phase transition at any finite temperature, the two-dimensional Ising model does undergo such a transition.} Near unstable fixed points, critical phenomena can be discussed. The critical phenomena and scaling laws near this fictitious fixed point are studied in Sec.~\ref{sec:Critical}.

Let us discuss the physical meaning of the RG transformation in the DNA. As discussed earlier, the introduction of dynamic neurons results in additional terms, $\lambda_m H_m$, which reflect the constraints,
\begin{equation}
   H_m= \left[h^{(m+1)} - \sigma\left(W^{(m)} h^{(m)}+b^{(m)} \right) \right]^2 = 0,\label{eq:const}
\end{equation}
equivalent to Eq.~\eqref{eq: recursive relation}. This constraint explains the relationship between the nearest degrees of freedom. After the RG transformation, the degrees of freedom in the $m$-th layer are integrated out, and the renormalized Hamiltonian $H_m'$ is composed of the nearest degrees of freedom.\footnote{Since the $m$-th layer is integrated out, the $(m-1)$-th and $(m+1)$-th layers become the nearest layers.} As Eq.~\eqref{eq:const} reflects the constraints between the nearest degrees of freedom, the vanishing conditions of the renormalized Hamiltonian  
\begin{equation}
 H_m' = 0   
\end{equation}
can be interpreted as new constraints between the nearest degrees of freedom in the renormalized system. As the depth of the DNN decreases by applying the RG transformation, new constraints for the system with reduced depth emerge. Therefore, the RG transformation suggests statistically equivalent constraints for the system with reduced depth.
Theoretically, by applying these new constraints to the partition function, the introduced dynamic neurons can be integrated out. Subsequently, a renormalized DNN system without dynamic neurons can be obtained. In general, after applying the RG transformation, the constraints often become more complex, leading to multiple solutions or analytically unsolvable equations. However, even if we cannot solve them explicitly, identifying these constraints can aid in understanding the structure of DNNs. In this sense, the DNA is a valuable tool for comprehending the nature of DNNs. This argument can be generalized without relying on the simplifying assumptions (i)–(iii).

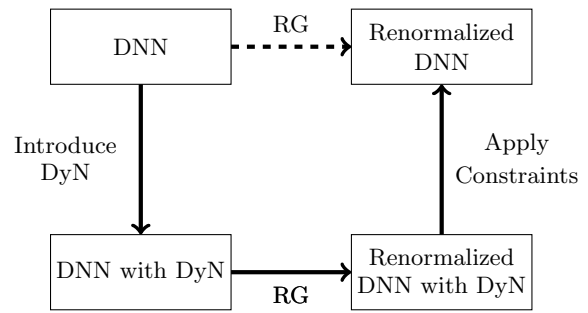
\begin{figure}[b]
    \centering
\begin{tikzpicture}
    \draw (-3.2,2) rectangle (-1+.2,1);
    \node at (-2,3/2) {DNN};
    \draw (-3.2,-2) rectangle (-1+.2,-1);
    \node at (-2,-3/2) {DNA};
    \draw (3.2,2) rectangle (1-.2,1);
    \node at (2,3/2+.2) {Renormalized};
    \node at (2,3/2-.2) {DNN};
    \node at (2,-3/2+.2) {Renormalized};
    \node at (2,-3/2-.2) {DNA};
    \draw (3.2,-2) rectangle (1-.2,-1);
    \draw[->,ultra thick, dashed] (-1+.2,3/2) -- (1-.2,3/2);
    \draw[->,ultra thick] (-1+.2,-3/2) -- (1-.2,-3/2);
    \draw[->,ultra thick] (-2,1) -- (-2,-1);
    \draw[->,ultra thick] (2,-1) -- (2,1);
    \node at (0,2-.2) {RG};
    \node at (0,-2+.2) {RG};
    \node at (-3,+.4) {Introduce};
    \node at (-3,0) {Dynamic};
    \node at (-3,-.37) {Neurons};
    \node at (3.4,+.4) {Apply};
    \node at (3.4,0) {Renormalized};
    \node at (3.4,-.37) {Constraints};
\end{tikzpicture}
\caption{This diagram illustrates how to develop a renormalized DNN using the DNA. Although not a rigorous explanation, it effectively conveys the essential concept. By introducing dynamic neurons, we obtain a new description of the DNN, and from the advantages of this new description, we derive a renormalized DNN with DNA. Finally, by applying renormalized constraints, the dynamic neurons can be integrated out, resulting in a renormalized DNN.}
\label{fig:renormalized deep learning}
\end{figure}

The discussions in this section are schematically summarized in Fig.~\ref{fig:renormalized deep learning}. Because the RG transformation is a powerful technique in statistical physics, we aim to use it to analyze DNNs. However, directly applying the RG transformation to DNNs is challenging, so we employ the DNA. Due to the translational symmetry that emerges from the DNA, the RG transformation becomes much easier to apply. When the DNN undergoes an RG transformation using the DNA, it suggests new constraints for the system with reduced depth. By applying these constraints, one can obtain a renormalized DNN system.

Before concluding this section, let us discuss how to relax the assumptions we introduced earlier in this section for the sake of demonstration. Using the decimation-based RG transformation may not be suitable when relaxing these assumptions. If assumption (i) is relaxed by increasing the sizes of $L$ and $N$, decimation can still be applied, but the number of couplings increases, making it challenging to analytically determine the fixed points. If assumption (ii) is generalized to allow different activation functions, assumption (iii) concerning the available values of the dynamic neuron must also be extended accordingly. If assumption (iii) is relaxed, expanding the available values of synaptic weights, biases, and dynamic neurons could result in an infinite number of possible values. In this case, polynomial expansion in decimation would no longer be feasible.

Nevertheless, these challenges arise from the difficulty of applying decimation, not from the limitations of the DNA itself. DNA can be applied regardless of assumptions (i), (ii), and (iii). Referring to the equations we presented in the previous section, extending the DNA to more general scenarios is straightforward. Statistical physics offers numerous RG transformation methods that do not rely on decimation. Applying these methods could enable further research on DNNs based on the DNA without these assumptions.

\section{Critical Phenomena}
\label{sec:Critical}

Critical phenomena are dramatic behaviors that occur near critical points and are crucial in understanding phase transition physics. One key feature of critical phenomena is the scaling law, where physical quantities exhibit power-law behavior. A remarkable aspect of critical phenomena is universality, which means that many systems can exhibit the same scaling laws despite microscopic differences, provided they share key characteristics.

To make predictions about the scaling properties of general deep learning, we further study the fictitious fixed point from the example in the previous section, even though it is not physical. By considering the linear approximation of the RG transformation near the critical point, we can analyze the critical properties of the system:
\begin{equation}\begin{split}
\vec{K}'-\vec{K}^*&=\mathcal{R}(\vec{K})-\mathcal{R}(\vec{K}^*)\\
   &\approx\frac{d\mathcal{R}}{dK}\bigg{|}_{K=K^*}(K-K^*),
\end{split}\end{equation}
where $\vec{K}=(e^A,e^B,e^C)$, $\vec{K}'=(e^{A'},e^{B'},e^{C'})$, $\vec{K}^*=(3,-2/3,-1/3)$, and $\mathcal{R}(\vec{K})=\vec{K}'$ denotes the function of RG transformation. Then, near $\vec{K}^*$, the RG transformation can be approximated as follows:
\begin{equation}\begin{split}
    \begin{bmatrix}
        e^{A'}-3\\
        e^{B'}+2/3\\
        e^{C'}+1/3
    \end{bmatrix}
    &\approx \begin{bmatrix}
        2 & 0 & 0 \\
        0 & 2 & 0\\
        0 & 0 & 2
    \end{bmatrix}
    \begin{bmatrix}
        e^{A}-3 \\
        e^{B}+2/3 \\
        e^{C}+1/3
    \end{bmatrix}\\
    &=\begin{bmatrix}
        p^{y_A} (e^{A}-3)\\
        p^{y_B} (e^{B}+2/3)\\
        p^{y_C} (e^{C}+1/3)
    \end{bmatrix}
    , \label{eq:linear}
\end{split}\end{equation}
where $p = 2$ denotes the ratio between the depth of the original and renormalized systems.\footnote{In general, by taking the continuum limit of the system, the parameter $p$ can be extended to real numbers. This enables the definition of an infinitesimal RG transformation with parameter $p=1+\epsilon$, where $\epsilon$ is an infinitesimally small quantity.} The values $y_A = 1$, $y_B = 1$, and $y_C = 1$ determine the universality class of the system. The values may seem trivial here, but this is due to the simplicity of the system we are considering.

From Eq.~\eqref{eq:linear}, we can find the scaling laws of observables near the fixed point. As an example, let us introduce the scaling of the correlation function. For simplicity, we explore the scaling laws under the conditions $B=B^*$ and $C=C^*$, though this argument can easily be generalized.

The correlation function of neurons is defined as
\begin{equation}\begin{split}
    G_h&(h^{(m)},h^{(n)}, A-A^*)\\& =\langle h^{(m)} h^{(n)}\rangle_{H}-\langle h^{(m)} \rangle_{H}\langle h^{(n)}\rangle_{H}\\ 
    & =G_h(m-n, A-A^*),
\end{split}\end{equation}
where $\langle \cdot \rangle_H$ denotes ensemble average with respect to the Hamiltonian $H$.\footnote{We can also define correlation functions for different variables, such as $G_\sigma(\sigma^{(m)},\sigma^{(n)}, A-A^*)$.} Note that the last equality holds due to translational invariance. As stated earlier, translational invariance was not readily apparent in the original cost function. This implies that our approach reveals the underlying correlating properties of deep learning.

The correlation length $\xi$ is defined from the behavior of the correlation function, which means that
\begin{equation}
    G_h(m-n,A-A^*)\propto e^{-|m-n|/\xi}
\label{eq: def correlation length}\end{equation}
for sufficiently large $|m - n|$.

After the RG transformation, the difference in depth between two neurons decreases by the ratio $p$. Since we are calculating the correlation function of neurons $h$, the coupling constants associated with the neurons, $B$, affect the scaling of the correlation function. Furthermore, the dimension of the system $d$, along with the degree of freedom we are reducing, affects the RG transformation. In our example, we consider the RG transformation with respect to depth, so $d = 1$.

In general, the following scaling behavior can be found near the critical point:
\begin{equation}
G_h\hspace{-1mm}\left(\frac{m-n}p, A'-A^*\right) \hspace{-1mm}= p^{2d-2y_B} G_h(m-n, A-A^*).
\label{eq: scaling behavior of correlation}
\end{equation}
Equation~\eqref{eq: scaling behavior of correlation} is the core of the scaling law for the correlation function and correlation length.

Considering $p|A-A^*|=A_0$ for some positive constant $A_0$, and substituting $p=(|A-A^*|/A_0)^{-1/y_A}$, one can conclude that
\begin{equation}
    G_h({m-n}, A-A^*)\propto \Psi\left((m-n)\left|\frac{A-A^*}{A_0}\right|^{1/y_A}\right)
\end{equation}
for some function $\Psi$. For large $m - n$, we expect the correlation function to follow Eq.~\eqref{eq: def correlation length}, so we find the scaling law of the correlation length:
\begin{equation}\xi\propto |A-A^*|^{-1/y_A}. \end{equation}
 Now, considering that $|m-n|/p=r_0$ for some fixed distance $r_0$, and repeating the same argument, one can conclude that
\begin{equation}G_h(m-n)\propto |m-n|^{-2d+2y_B}\end{equation}
near the critical point.\footnote{{Here, we define ``distance'' as the number of layers separating two neurons. In more complex deep neural network architectures, this concept may need a more refined definition.}} 
These scaling laws are frequently observed in various examples of critical phenomena in statistical physics. Further research from the statistical physics perspective could provide deeper insights into these phenomena.

\begin{figure}
    \centering
    \includegraphics[width=0.85\linewidth]{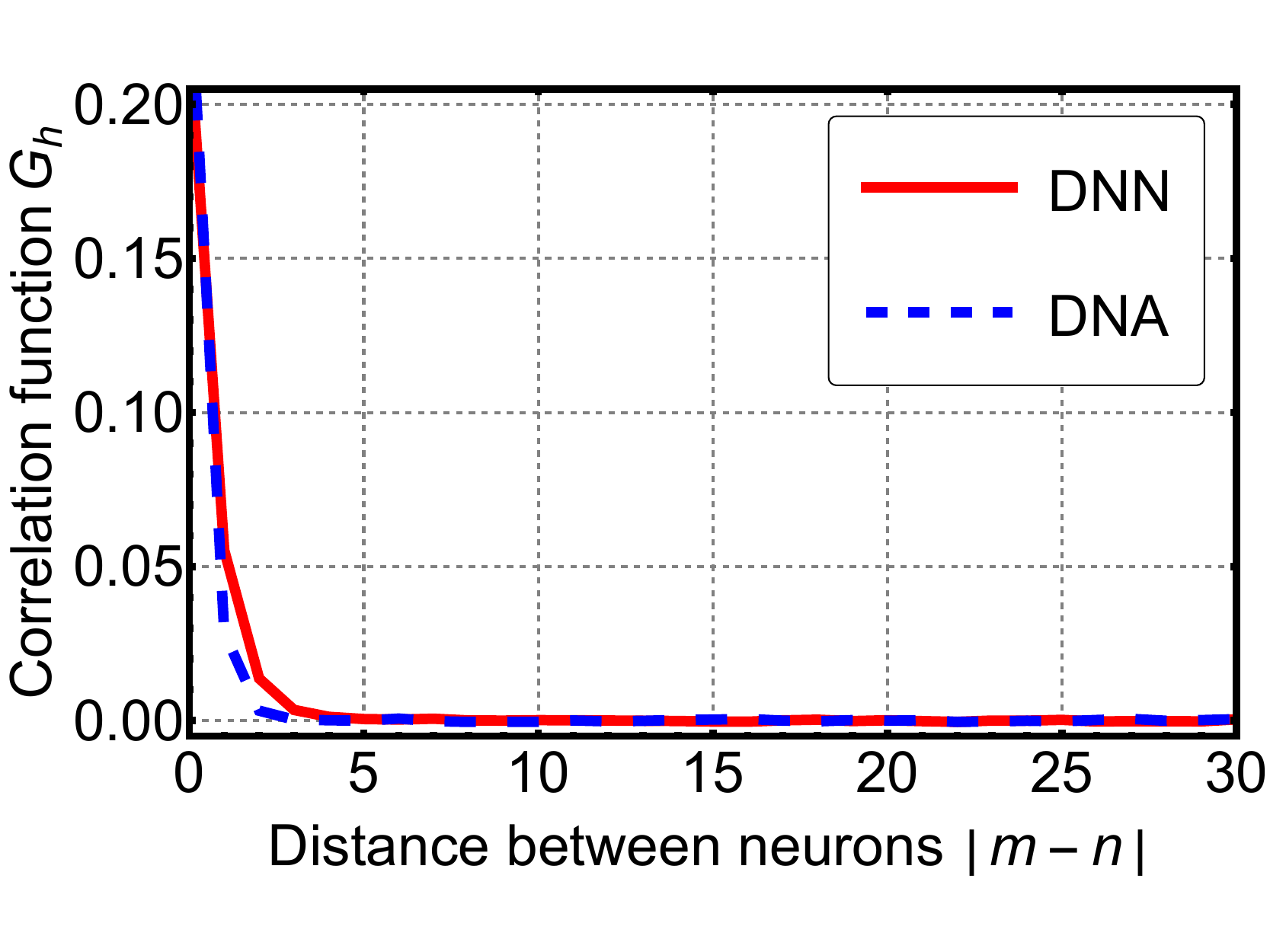}
    \caption{Results of a Monte Carlo simulation for a simple neural network (red solid) and its dynamic neuron counterpart (blue dashed). We computed the time-averaged correlation function of neuron activations as a function of spatial separation, averaged over all positions. The two systems exhibit a close resemblance. } 
    \label{fig:Correlation}
\end{figure}

Equation~\eqref{eq:modified pf} implies that, in the limit of large $\beta_v$, the DNA becomes equivalent to the DNN. To examine this correspondence, we performed a Monte Carlo simulation of the simplified neural network introduced in the previous section. The network depth was set to $10^4$, and we used $\beta = \beta_v = 1$. As shown in Fig.~\ref{fig:Correlation}, the time-averaged correlation functions of both systems exhibit a close resemblance. This observation suggests that, under suitable conditions, the DNA can serve as a reasonable theoretical approximation of the DNN. Owing to its analytical tractability, the DNA provides a useful perspective for further investigations into the structure and dynamics of deep neural networks.

\section{Summary and Discussion}
\label{sec:Summary}

The statistical physical approach to deep learning has been explored in many studies, but there is no agreed-upon description. Additionally, the dynamics of neurons have only been discussed in specific models, such as the Hopfield network and the Boltzmann machine, which differ fundamentally from typical deep learning systems \cite{hopfield1982neural, little1974existence, montufar2018restricted}.

Our study introduced the DNA. In this approach, neurons are treated as dynamic variables of the system. We transformed the original Hamiltonian, where the degrees of freedom are synaptic weights and biases, into a new Hamiltonian with additional degrees of freedom: dynamic neurons. The DNA with such a transformed system could be understood as a generalization of deep learning, where the limit of large $\beta_v$ corresponds to the typical DNN.

The new Hamiltonian simplifies the interactions and reveals translational invariance. This symmetry, induced by the DNA, is approximate because the first and last layers are ignored. However, as the number of layers increases, the relative effects of the first and last layers diminish, and the symmetry becomes better preserved. This observation could be related to a fundamental question about DNNs: why do they perform so well when their depth is large?  Our approach might offer insights into many fundamental questions about DNNs \cite{futurework}.

Additionally, the DNA highlights that the degrees of freedom—dynamic neurons, synaptic weights, and synaptic biases—interact primarily with their nearest neighbors. This interaction is intuitive, as neurons in a network communicate only with adjacent neurons to produce the final output. Consequently, the entire system can be understood by examining its individual components and their interactions.

We discussed the RG transformation and critical phenomena using a simple example of a DNN with dynamic neurons. Although it was a straightforward case, we demonstrated that well-known statistical approaches can be easily applied to DNNs using the DNA. RG transformations and critical phenomena are deeply studied topics in statistical physics with many physical implications. Finding connections between these implications and the properties of DNNs is very intriguing and promises rich outcomes.

Although our study explains the statistical property of the specific example, it should be emphasized that the DNA can be applied to any DNN with arbitrary width, depth, and the number of data points. Many studies on DNNs concentrate on factors such as network size and the amount of training data. We anticipate that our DNA can become a useful tool to investigate those subjects.

In the present paper, given the limited theoretical development surrounding inference, we have focused on building a model that analytically explains the statistical properties of the training process. Addressing inference from a theoretical perspective remains challenging, as it involves understanding how neural networks generalize to unseen data sets. Future work should therefore develop theoretical tools for inference that would enable investigations into phenomena such as the neural scaling law \cite{liao2020random, kaplan2020scaling, caballero2022broken, bahri2021explaining, sorscher2022beyond, hastie2022surprises, ziyin2022exact, bordelon2024dynamical, ma2024neural, du2024understanding} and its relevance to our theory. These advances will help bridge the gap between the statistical properties of training and predictive performance in practical applications.

\begin{acknowledgments}
This work was supported in part by the National Research Foundation of Korea (Grant No. RS-2024-00352537).
Author names in this work are in alphabetical order.
\end{acknowledgments}

\appendix

\section{Technical Application}
\label{Appendix:Technical}

This appendix delves into the technical implementation of the dynamic neuron system. Since the DNA modifies the Hamiltonian and introduces new degrees of freedom, referred to as ``dynamic neurons," it is crucial to analyze both the altered system dynamics and the behavior of these newly introduced components.

The new Hamiltonian described in Eq.~\eqref{eq: modified Hamiltonian} emerges from the dynamic neuron framework and governs the system's behavior. Consequently, this introduces additional equations of motion for dynamic neurons and modifies the equations of motion for synaptic weights accordingly. To implement such a system, we can utilize the gradient descent method, as outlined in Eq.~\eqref{eq:gradient_descent}, in the following manner.
\begin{equation}\label{eq: technical realization}
    \begin{split}
        &\Delta{W}_{ij}^{(m)}=-\eta\frac{\partial H_{\text{new}}}{\partial W_{ij}^{(m)}}\\
        &\Delta{b}_{i}^{(m)}=-\eta\frac{\partial H_{\text{new}}}{\partial b_{i}^{(m)}}\\
        &\Delta{h}_{i}^{(m)[l]}=-\eta\frac{\partial H_{\text{new}}}{\partial{h}_{i}^{(m)[l]}}
    \end{split}
\end{equation}
We need to implement changes for the dynamic neurons for each data represented by $l$.
Under the lens of our theoretical framework, we anticipate that when $\beta_v$ is large enough the conventional DNN and the dynamic neuronal DNN will exhibit only minor differences in their statistical properties over the same training epochs.

However, it seems that constructing a DNN using the DNA is impractical at this point.
In this framework, the number of dynamic neurons introduced is $LMN$, corresponding to the product of the number of data points, the width of the network and its depth. This implies that additional $LMN$ equations of motion must be solved to implement the system. Given the large datasets typically used in training, this approach appears computationally infeasible for practical applications.

Nonetheless, we remain open to the potential technical applications of this approach. Despite its current inefficiency, the development of better optimization strategies could enable the practical realization of the dynamic neuron framework. In many deep learning implementations, Eq.~\eqref{eq:gradient_descent} are not solved simultaneously; instead, more efficient methods like backpropagation are commonly used for training. A similar method could be designed specifically for DNNs based on the DNA. The inherent homogeneity in the equations of motion, as highlighted in our main text, could serve as a basis for innovative optimization techniques. Moreover, since interactions in the DNA are restricted to nearest neighbors, the right-hand sides of Eq.~\eqref{eq: technical realization} are expected to be simpler than those of Eq.~\eqref{eq:gradient_descent}. With further research into such optimization methods, we expect that the practical realization of the DNA could become feasible.

\section{Spin Glass}
\label{Appendix:SpinGlass}

In this appendix, we illustrate how the new Hamiltonian in the DNA replicates key characteristics of spin glass models, a class of statistical models renowned for their complex energy landscapes and disordered states. While there are several formulations of spin glass models, we focus on the Edwards-Anderson model, a canonical example widely used in statistical physics \cite{edwards1975theory, nishimori2001statistical}:
\begin{equation}
H = \sum_{i,j} J_{ij} s_i s_j + \sum_i 
 K_i s_i,
\end{equation}
where $s_i$ are spin degrees of freedom, which take values of $1$ or $-1$, and the coupling constants between spins $J_{ij}$ are typically assumed to follow independent probability distributions, such as Gaussian or Bernoulli distributions. $K_i$ is the coupling to magnetic fields, which depends on an external magnetic field.

We apply certain conditions and assumptions to the new Hamiltonian presented in Eq.~\eqref{eq: modified Hamiltonian}. We suppose the activation function is the sigmoid function, so $|\sigma|$ and $|h|$ are likely smaller than $1$. Also, we introduce the regularization terms to restrict $W$ and $b$ from becoming too large. Furthermore, we expect that the linear order expansion of the sigmoid function is a valid approximation.

The new Hamiltonian, without the constant term, is then given as:
\begin{equation}
\begin{split}
    H=\sum_{i, j, l, m} & \lambda_m W_{ij}^{(m)} h_i^{(m+1)[l]}h_j^{(m)[l]}\\& +\sum_{i,l,m}\lambda_m b_i^{(m)}h_i^{(m+1)[l]}\\ & + \sum_{i,j,m} c_W (W_{ij}^{(m)})^2 + \sum_{i,m} c_b (b_i^{(m)})^2.
\end{split}
\end{equation}
where the last two terms are regularization terms. These regularization terms provide the Gaussian probability distribution for $W$ and $b$ and result in the Gaussian distributed coupling constants for the first two terms in the Hamiltonian. Then the first two terms of Hamiltonian are similar to the spin glass model with a space-varying external magnetic field with degrees of freedom $h$ and coupling constants determined by $W$ and $b$.\footnote{One should note that despite these similarities, there are differences between spin glass models and the new Hamiltonian: (1) The couplings can vary in time, and (2) The spin degrees of freedom can take arbitrary values between $-1$ and $1$.}

\newpage

\bibliography{ref.bib}

\end{document}